\documentclass[runningheads]{llncs}
\usepackage{graphicx}
\usepackage{url}

\usepackage{graphicx}
\usepackage{float}
\usepackage{multirow}
\usepackage{booktabs}
\usepackage{amsmath}
\usepackage{enumitem}
\usepackage{longtable}

\graphicspath{ {./images/} }
\begin{document}
\title{Perception and Localization of Macular Degeneration Applying Convolutional Neural Network, ResNet and Grad-CAM}
\titlerunning{Classify and Localize Macular Degeneration using CNN, ResNet, Gradcam}
%
\author{}
\institute{}
\author{Tahmim Hossain\inst{1,2}  \and
Sagor Chandro Bakchy\inst{1,3}
}
%
%
\institute{Rajshahi University of Engineering \& Technology, Rajshahi, Bangladesh 
\and \email{tahmimhossain.official@gmail.com}\\
\and \email{sagorchandro.10@gmail.com}\\
}

\maketitle              
\begin{abstract}
A well-known retinal disease that feels blurry visions to the affected patients is Macular Degeneration. This research is based on classifying the healthy and macular degeneration fundus with localizing the affected region of the fundus. A CNN architecture and CNN with ResNet architecture (ResNet50, ResNet50v2, ResNet101, ResNet101v2, ResNet152, ResNet152v2) as the backbone are used to classify the two types of fundus. The data are split into three categories including (a) Training set is 90\% and Testing set is 10\% (b) Training set is 80\% and Testing set is 20\%, (c) Training set is 50\% and Testing set is 50\%. After the training, the best model has been selected from the evaluation metrics. Among the models, CNN with backbone of ResNet50 performs best which gives the training accuracy of 98.7\% for 90\% train and 10\% test data split. With this model, we have performed the Grad-CAM visualization to get the region of affected area of fundus.

\end{abstract}
\section{Introduction}

Macular Degeneration is a condition in which the central vision gradually deteriorates caused by a chronic retinal illness that affects the macular portion of the retina~\cite{ref1,ref3}. By 2040, the world's population of persons with macular degeneration disease is predicted to reach over 300 million~\cite{ref2,ref4}. Patients may feel blurring or distorted visuals in their central visual field known as metamorphopsia. For worsened conditions, the center field may become obliterated, and central vision loss may occur~\cite{ref5}. Macular degeneration affects around 5\% of the population under the age of 65, according to several epidemiological studies~\cite{ref6}. However, it is more common in older people, with more than 35\% of those over the age of 75 being impacted in some way~\cite{ref7}. Retina’s central, posterior part is known as macula. High-resolution visual occurs for the photo-receptors’ densest concentration~\cite{ref8}. 

Early detection can be great to prevent the macular degeneration disease by giving the proper treatment to the patient who is affected. This can help the medical practitioner to take the necessary steps as soon as the disease is detected. It may help a person to not lose his vision permanently. Automation can be greatly helpful to limit the spreading of the disease inside the eye. As the Macular Degeneration can be formed in ‘Dry’ or ‘Wet’~\cite{ref9}. Dry MD has three stages known as early, intermediate, and late and any stage of dry MD can result in wet Macular Degeneration~\cite{ref10}. So, from the early or intermediate stage of the disease can be determined it will help the person to get rid away from the disease easily and give treatment to the affected region which will get from the Grad-CAM visualization. 
Firstly two types of fundus images have been classified by analyzing the dataset~\cite{ref13}. The motivation of this study is to get the perfect CNN model and Transfer learning model (ResNet Models) added to the CNN model by tuning the hyperparameters to classify the Macular Degeneration affected fundus and healthy fundus and localize the affected region using Grad-CAM visualization. A Convolutional Neural Network model and then CNN model with backbone of transfer learning models including ResNet50, ResNet50v2, ResNet101, ResNet101v2, ResNet152, ResNet152v2~\cite{ref11,ref12} is used to do a benchmark. The hyperparameters include optimizers, epoch, batch size, shuffle, data train, and test splitting. The data has been split into ways that include 90\% of training set 10\% of testing set, 80\% of training set 20\% of testing set, and 50\% of training set 50\% of testing set. By evaluating the performances of the models, the top model is CNN with the backbone of ResNet50. After that Grad-CAM is applied to visualize the important region of the input images.

\section{Related Works} 

In this section, we discuss previous literature for MD classification and data augmentation since this paper contributes to automated macular degeneration classification in terms deep learning and machine learning approaches and data augmentation methodologies. It is important to note that we concentrate on deep learning methodologies.

Wang et al.~\cite{ref14} use a CNN capable of four-class prediction with two data augmentation approaches. Using ResNet18 as its backbone, the authors create a multi-modal CNN that predicts a specific class based on OCT and fundus types of pictures acquired from patient eyes. The dataset originally contains 1,099 color fundus pictures from 1,099 eyes. Given a relatively modest real-world dataset, they can create a large number of multimodal training examples by combining the suggested CAM-conditioned picture synthesis and loose pairing.

Burlina et al. for example, attempt to solve AMD classification challenge based on 2 class in~\cite{ref15} by classifying fundus pictures of individuals with non-existence or early-stage AMD affected patient with intermediate or advanced stage AMD affected patients. They solve referable AMD classification issue using more than 130000 photos from the AREDS dataset using the AlexNet DCNN model. They next discuss preprocessed fundus images by identifying the retina's outer margins, cropping images to the square inscribed inside the retinal boundary, and scaling the square to fit AlexNet or OverFeat DCNNs' expected input size.

Keel et al.~\cite{ref16} use Inception v3 architecture to create and train a deep learning algorithm (DLA) for the detection of neovascular age-related macular degeneration using color fundus photography, using three deep learning-based models. Scaling, eliminating the local space average color, compressing the image to a 299x299 matrix, and data augmentation for normalization are all used to preprocess the image in their work.

For performing segmentation, feature extraction and regression tasks, Perdomo Charry et al.~\cite{ref19} propose a model containing 4 main stages with 3 deep learning-based architectures t designed with a big number of layers. In this research, they used The RIM-ONE v3 dataset with 159 eye color fundus photos and The REFUGE challenge dataset with 1200 retinal fundus images. They also extract characteristics from ocular fundus pictures using a ResNet50 pre-trained D-CNN. They achieve automated extraction of polar morphometric features from eye fundus pictures using a modified D-CNN depended on the architecture of VGG-19 and VGG-16.
  
\section{Dataset} 
The dataset~\cite{ref13} is taken from Kaggle which is a competition named VietAI Advance Course - Retinal Disease Detection.  The dataset has been taken from Cao Thang Eye Hospital (CTEH) hospital of Vietnam. The collection comprises 3,285 CTEH pictures (3.210 abnormal and 75 normal) and 500 Messidor and EYEPACS normal images. Opacity, diabetic retinopathy, glaucoma, macular edema, macular degeneration, and retinal vascular occlusion are among the anomalies. Fundus imaging gives doctors a picture of the inside of patient's eye. Fundus imaging can detect a variety of eye illnesses, including diabetic retinopathy, glaucoma, and macular degeneration. In this research work, I have extracted the macular degeneration fundus images and healthy fundus images by exploring the dataset. There were 299 macular degeneration fundus images and 525 normal fundus images which was an unbalanced dataset. Then I downsampled both classes into 299 macular degeneration images and 299 normal fundus images for a total of 598 fundus images.

\section{Methodology}
\begin{figure*}[t]
\centering
  \includegraphics[width=1\linewidth]{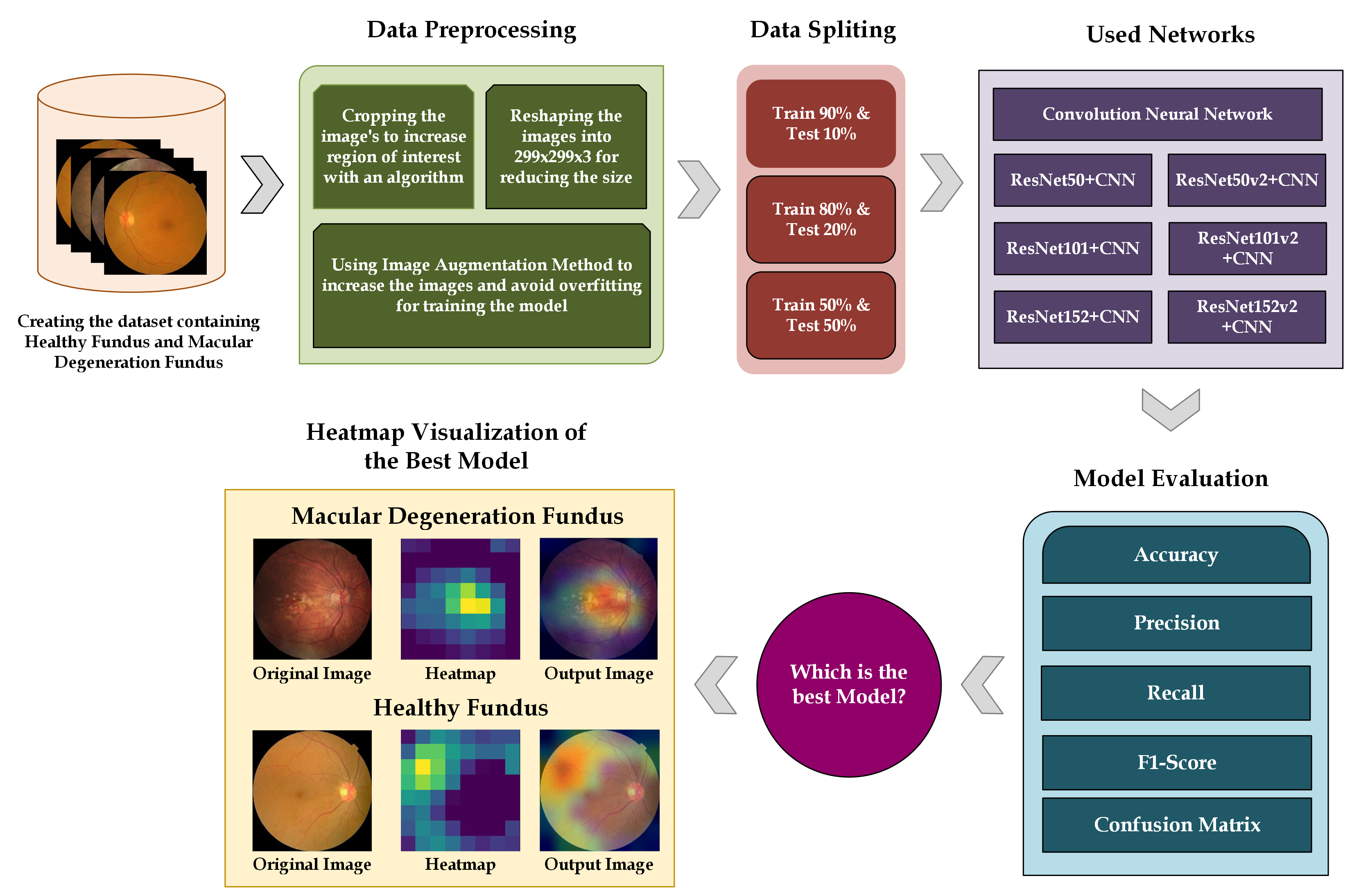}
\caption{Working flow diagram  of our whole research work}
\label{fig1}       
\end{figure*}

Firstly, the healthy fundus and Macular Degeneration fundus has been extracted from the dataset and the dataset was made balanced. After that preprocessing techniques including image cropping, reshaping is used. We have used seven models to benchmark the dataset and from the seven models the best model is selected. The selection process of the best model has been done from the evaluation metrics. Then after selecting the best model we have used Grad-CAM to localize the affected area, Figure ~\ref{fig1} shows the summarized methodology that is used throughout this paper.

\section{Dataset Preprocessing}
\subsection{Image Cropping and Reshaping}
Initially the images were consisted of many black areas which may results the accuracy of the model too low. That’s why the images have been cropped according to the size of fundus with small amount of black portion. The below figures show the before and after version of cropping the images. In the reshaping technique, the initial size of the images was huge. Such as for single image it was around 2-3MB. So, it has been resized to 299x299x3. It has been helped to train the model faster than before.

\subsection{Image Augmentation}
As the limitation of the data the image augmentation has been performed. Horizontal Flip, Vertical Flip and Random Rotation has been used to get an optimized training from the model. The horizontal flip means the right side will shift to the left side or vice versa and the vertical flip means the bottom part will shift to the top part or vice versa. The random rotation means the images will rotate in a random angle and generate random rotated images. Below you can see the examples of the Data augmentation.

\section{Convolutional Neural Network Model}
\begin{figure*}[t]
\centering
  \includegraphics[width=1\linewidth]{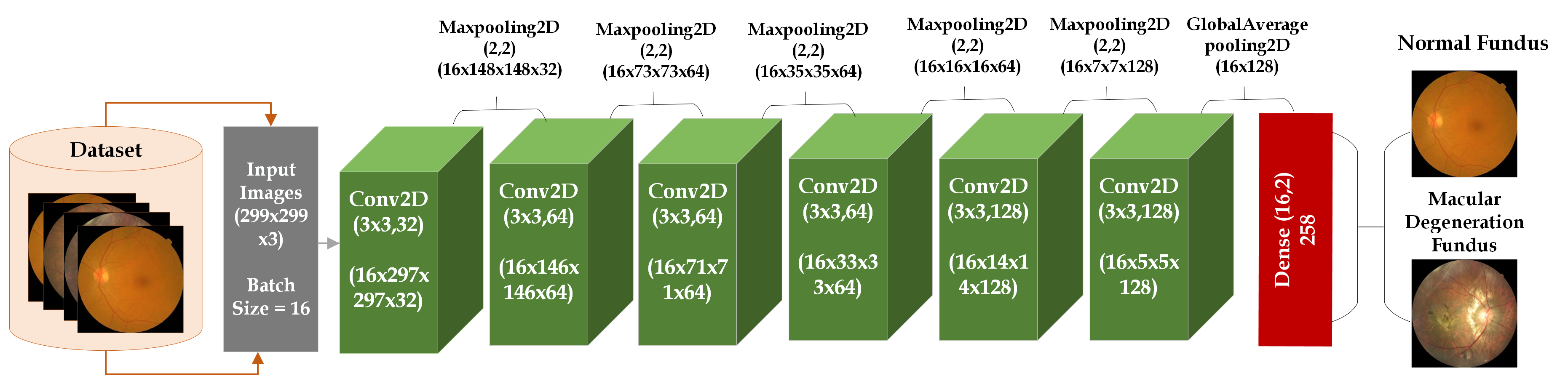}
\caption{Schematic Diagram of CNN Model}
\label{fig2}       
\end{figure*}

The input image size is 299x299x3 with the batch size of 16. The used convolution has six layers with kernel size of 32, 64, 64, 64, 128, 128 respectively. Each layer has got MaxPooling2D Later with pool size of (2,2). The dense layer and last Conv2D layer is connected with a Global Average Pooling 2D layer with default pool size. From the dense layer the final output has been shown. Figure~\ref{fig2} shows the schematic diagram of the convolutional neural network model.

\section{RestNet+CNN Architecture}
\begin{figure*}[t]
\centering
  \includegraphics[width=1\linewidth]{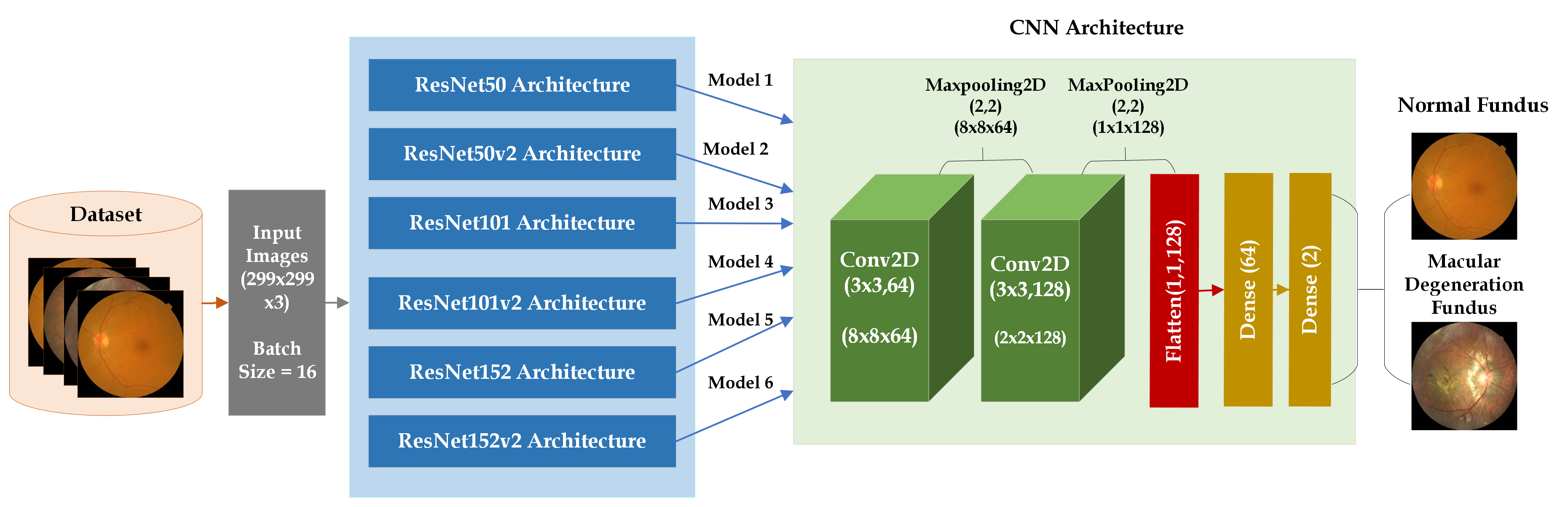}
\caption{Schematic Diagram of Six ResNet Models with CNN Models}
\label{fig3}       
\end{figure*}

Research named 'Deep Residual Learning for Image Recognition,' Kaiming He et al. bought forward an innovative neural network in 2015~\cite{ref11}. Six models have been created with a CNN and the backbone of ResNet50, ResNet50v2, ResNet101, ResNet101v2, ResNet152, ResNet152v2. Figure 5 shows the schematic diagram of six models.  The models have been created with the help of Keras API. Detailed description of each model is illustrated in Figure~\ref{fig3}.

\subsection{ResNet50+CNN (Model 1)}
With about 23 million trainable parameters, the ResNet50 model is consisted of 48 CNN layer with one Max Pool and one Average Pool layer in total of 50 neural network layer~\cite{ref11}. ResNet50 is used as the backbone of two Conv2D layer is also added with the 2-max pooling 2D layer and two dense layer which makes in total of 24.85M parameters.

\subsection{ResNet50v2+CNN (Model 2)}
ResNet50V2, as an updated version of ResNet50, outperforms ResNet50 and ResNet101 on the ImageNet dataset~\cite{ref12}. The propagation formulation of the links between blocks was altered in ResNet50V2. ResNet50v2 is used as the backbone of two Conv2D layer is also added with the 2-max pooling 2D layer and two dense layer which makes in total of 24.85M parameters. 

\subsection{ResNet101+CNN (Model 3)}
A convolutional neural network consisting of 101 layers is known as ResNet-101~\cite{ref21}. ResNet101 is used as the backbone of two Conv2D layer is also added with the 2-max pooling 2D layer and two dense layer which makes in total of 43.92M parameters.

\subsection{ResNet101v2+CNN (Model 4)}
ResNet-101v2 is a modified version of ResNet101 with 101 layers~\cite{ref11}. ResNet101 is used as the backbone of two Conv2D layer is also added with the 2-max pooling 2D layer and two dense layer which makes in total of 43.88M parameters.

\subsection{ResNet152+CNN (Model 5)}
All convolutional layers in ResNet-152 use the same convolutional window of size 3x3, while the number of filters rises with network depth, from 64 to 2048. There is only one max-pooling layer which is 2 is applied after the first layer. To replace completely connected layers, the average pooling layer is employed. At addition, in this layer, a softmax activation function is used to calculate the likelihood that the input belongs to each class~\cite{ref21}. ResNet152 is used as the backbone of two Conv2D layer is also added with the 2-max pooling 2D layer and two dense layer which makes in total of 59.63M parameters.

\subsection{ResNet152v2+CNN (Model 6)}
To replace completely connected layers, the average pooling layer is used. At addition, in this layer, a softmax activation function is used to calculate the likelihood that the input belongs to each class~\cite{ref22}. ResNet152v2 is used as the backbone of two Conv2D layer is also added with the 2-max pooling 2D layer and two dense layer which makes in total of 63.35M parameters.

\section{Performance Evaluation of Models}
In this research, seven models have been used to classify macular degeneration fundus and healthy fundus. According to evaluate those models the ultimate decision has been made regarding the score of accuracy, precision, sensitivity, F1 score and visualizing the confusion matrix. These scores are generally defined from test set of the dataset. Accuracy gives the idea that how much output is predicting right. Precision score defines that how exactly the fundus diseases are predicting. The sensitivity refers to the percentage of relevant diseases that can be matched up. F1-score defines the model accuracy considering the precision and recall. In the confusion matrix there are four blocks for binary classification. The four blocks are defined as true positive ($T_P$), true negative ($T_N$), false positive ($F_P$) and false negative ($F_N$). True positive and true negative are the diagonal elements of a confusion matrix. Calculating a confusion matrix can assist in determining where a classification model is succeeding and where it is failing. The equations that are used to evaluate the networks are given below:

\begin{equation}
    Accuracy = \frac{T_P + T_N}{T_P + T_N + F_N + T_N}\\
\end{equation}
\begin{equation}
    Precision = \frac{T_P}{T_P + F_P}\\
\end{equation}
\begin{equation}
    Sensitivity = \frac{T_P}{T_P+F_N}\\
\end{equation}
\begin{equation}
    F1-Score = \frac{2*(Sensitivity * Precision)}{Sensitivity + Precision}
\end{equation}

\section{Method of Visualization}
Gradient-weighted Class Activation Mapping (Grad-CAM) employs any target concept's gradients to produce a crude localization map for concept prediction that highlights key areas in the image~\cite{ref20}. Here after the analysis of all performance metrics heatmap is generated and applied to the fundus image to localize the important region of the image.

\section{Experimental Analysis}
\subsection{Performance metrics Analysis}
In this research, performance metrics has been considered to assess the best model. Performance metrics includes training accuracy, precision, sensitivity, and F1 Score. The dataset’s image size is 299x299x3 with the batch number of 16. To train the model the dataset has been divided into three ways. They are:
\begin{itemize}
    \item Training set is 90\% and Testing set is 10\% where for testing, it has used 37 of Macular Degeneration Fundus Images and 27 of Normal Fundus images.
    \item Training set is 80\% and Testing set is 20\% where for testing, it has used 70 of Macular Degeneration Fundus Images and 58 of Normal Fundus images.

\begin{longtable}{l|ccc}
\caption{Accuracy of all models}
\label{tab1}\\
\hline
\multirow{3}{*}{\textbf{Model}}  & \multicolumn{3}{c}{\textbf{Train Accuracy}}                                                                        \\ \cline{2-4} 
                                 & \multicolumn{3}{c}{\textbf{(Train Split\%)+(Test Split\%)}}                                                        \\ \cline{2-4} 
                                 & \multicolumn{1}{c|}{\textbf{90\%+10\%}}     & \multicolumn{1}{c|}{\textbf{80\%+20\%}}     & \textbf{50\%+50\%}     \\ \hline
\endfirsthead
\multicolumn{4}{c}%
{{\bfseries Table \thetable\ continued from previous page}} \\
\hline
\multirow{3}{*}{\textbf{Model}}  & \multicolumn{3}{c}{\textbf{Train Accuracy}}                                                                        \\ \cline{2-4} 
                                 & \multicolumn{3}{c}{\textbf{(Train Split\%)+(Test Split\%)}}                                                        \\ \cline{2-4} 
                                 & \multicolumn{1}{c|}{\textbf{90\%+10\%}}     & \multicolumn{1}{c|}{\textbf{80\%+20\%}}     & \textbf{50\%+50\%}     \\ \hline
\endhead
\multirow{2}{*}{CNN}             & \multicolumn{1}{c|}{\multirow{2}{*}{0.970}} & \multicolumn{1}{c|}{\multirow{2}{*}{0.966}} & \multirow{2}{*}{0.952} \\
                                 & \multicolumn{1}{c|}{}                       & \multicolumn{1}{c|}{}                       &                        \\ \hline
\multirow{2}{*}{ResNet50+CNN}    & \multicolumn{1}{c|}{\multirow{2}{*}{0.987}} & \multicolumn{1}{c|}{\multirow{2}{*}{0.976}} & \multirow{2}{*}{0.989} \\
                                 & \multicolumn{1}{c|}{}                       & \multicolumn{1}{c|}{}                       &                        \\ \hline
\multirow{2}{*}{ResNet50v2+CNN}  & \multicolumn{1}{c|}{\multirow{2}{*}{0.94}}  & \multicolumn{1}{c|}{\multirow{2}{*}{0.927}} & \multirow{2}{*}{0.945} \\
                                 & \multicolumn{1}{c|}{}                       & \multicolumn{1}{c|}{}                       &                        \\ \hline
\multirow{2}{*}{ResNet101+CNN}   & \multicolumn{1}{c|}{\multirow{2}{*}{0.986}} & \multicolumn{1}{c|}{\multirow{2}{*}{0.977}} & \multirow{2}{*}{0.98}  \\
                                 & \multicolumn{1}{c|}{}                       & \multicolumn{1}{c|}{}                       &                        \\ \hline
\multirow{2}{*}{ResNet101v2+CNN} & \multicolumn{1}{c|}{\multirow{2}{*}{0.919}} & \multicolumn{1}{c|}{\multirow{2}{*}{0.948}} & \multirow{2}{*}{0.925} \\
                                 & \multicolumn{1}{c|}{}                       & \multicolumn{1}{c|}{}                       &                        \\ \hline
\multirow{2}{*}{ResNet152+CNN}   & \multicolumn{1}{c|}{\multirow{2}{*}{0.983}} & \multicolumn{1}{c|}{\multirow{2}{*}{0.989}} & \multirow{2}{*}{0.97}  \\
                                 & \multicolumn{1}{c|}{}                       & \multicolumn{1}{c|}{}                       &                        \\ \hline
\multirow{2}{*}{ResNet152v2+CNN} & \multicolumn{1}{c|}{\multirow{2}{*}{0.859}} & \multicolumn{1}{c|}{\multirow{2}{*}{0.902}} & \multirow{2}{*}{0.935} \\
                                 & \multicolumn{1}{c|}{}                       & \multicolumn{1}{c|}{}                       &                        \\ \hline
\end{longtable}

\begin{longtable}{l|c|ccc|ccc|ccc}
\caption{Results of Precision, Sensitivity and F1-Score of all models}
\label{tab2}\\
\hline
\multirow{3}{*}{\textbf{Network}} &
  \multirow{3}{*}{\textbf{\begin{tabular}[c]{@{}c@{}}Fundus\\ Disease\end{tabular}}} &
  \multicolumn{3}{c|}{\textbf{Precision}} &
  \multicolumn{3}{c|}{\textbf{Sensitivity}} &
  \multicolumn{3}{c}{\textbf{F1-Score}} \\ \cline{3-11} 
 &
   &
  \multicolumn{3}{c|}{\textbf{(Train Spilt\%)}} &
  \multicolumn{3}{c|}{\textbf{(Train Spilt\%)}} &
  \multicolumn{3}{c}{\textbf{(Train Spilt\%)}} \\ \cline{3-11} 
 &
   &
  \multicolumn{1}{c|}{\textbf{90\%}} &
  \multicolumn{1}{c|}{\textbf{80\%}} &
  \textbf{50\%} &
  \multicolumn{1}{c|}{\textbf{90\%}} &
  \multicolumn{1}{c|}{\textbf{80\%}} &
  \textbf{50\%} &
  \multicolumn{1}{c|}{\textbf{90\%}} &
  \multicolumn{1}{c|}{\textbf{80\%}} &
  \textbf{50\%} \\ \hline
\endfirsthead
\multicolumn{11}{c}%
{{\bfseries Table \thetable\ continued from previous page}} \\
\hline
\multirow{3}{*}{\textbf{Network}} &
  \multirow{3}{*}{\textbf{\begin{tabular}[c]{@{}c@{}}Fundus\\ Disease\end{tabular}}} &
  \multicolumn{3}{c|}{\textbf{Precision}} &
  \multicolumn{3}{c|}{\textbf{Sensitivity}} &
  \multicolumn{3}{c}{\textbf{F1-Score}} \\ \cline{3-11} 
 &
   &
  \multicolumn{3}{c|}{\textbf{(Train Spilt\%)}} &
  \multicolumn{3}{c|}{\textbf{(Train Spilt\%)}} &
  \multicolumn{3}{c}{\textbf{(Train Spilt\%)}} \\ \cline{3-11} 
 &
   &
  \multicolumn{1}{c|}{\textbf{90\%}} &
  \multicolumn{1}{c|}{\textbf{80\%}} &
  \textbf{50\%} &
  \multicolumn{1}{c|}{\textbf{90\%}} &
  \multicolumn{1}{c|}{\textbf{80\%}} &
  \textbf{50\%} &
  \multicolumn{1}{c|}{\textbf{90\%}} &
  \multicolumn{1}{c|}{\textbf{80\%}} &
  \textbf{50\%} \\ \hline
\endhead
\multirow{2}{*}{CNN} &
  \begin{tabular}[c]{@{}c@{}}Healthy\\ Fundus\end{tabular} &
  \multicolumn{1}{c|}{1.00} &
  \multicolumn{1}{c|}{0.95} &
  0.9 &
  \multicolumn{1}{c|}{1.00} &
  \multicolumn{1}{c|}{0.98} &
  0.96 &
  \multicolumn{1}{c|}{1.00} &
  \multicolumn{1}{c|}{0.97} &
  0.93 \\ \cline{2-11} 
 &
  \begin{tabular}[c]{@{}c@{}}Macular\\ Degeneration\end{tabular} &
  \multicolumn{1}{c|}{1.00} &
  \multicolumn{1}{c|}{0.99} &
  0.96 &
  \multicolumn{1}{c|}{1.00} &
  \multicolumn{1}{c|}{0.96} &
  0.92 &
  \multicolumn{1}{c|}{1.00} &
  \multicolumn{1}{c|}{0.97} &
  0.94 \\ \hline
\multirow{2}{*}{\begin{tabular}[c]{@{}l@{}}ResNet50\\ +CNN\end{tabular}} &
  \begin{tabular}[c]{@{}c@{}}Healthy\\  Fundus\end{tabular} &
  \multicolumn{1}{c|}{1.00} &
  \multicolumn{1}{c|}{1.00} &
  1.00 &
  \multicolumn{1}{c|}{1.00} &
  \multicolumn{1}{c|}{0.98} &
  0.98 &
  \multicolumn{1}{c|}{1.00} &
  \multicolumn{1}{c|}{0.99} &
  0.99 \\ \cline{2-11} 
 &
  \begin{tabular}[c]{@{}c@{}}Macular\\ Degeneration\end{tabular} &
  \multicolumn{1}{c|}{1.00} &
  \multicolumn{1}{c|}{0.99} &
  0.98 &
  \multicolumn{1}{c|}{1.00} &
  \multicolumn{1}{c|}{1.00} &
  1.00 &
  \multicolumn{1}{c|}{1.00} &
  \multicolumn{1}{c|}{0.99} &
  0.99 \\ \hline
\multirow{2}{*}{\begin{tabular}[c]{@{}l@{}}ResNet50v2\\ +CNN\end{tabular}} &
  \begin{tabular}[c]{@{}c@{}}Healthy\\ Fundus\end{tabular} &
  \multicolumn{1}{c|}{1.00} &
  \multicolumn{1}{c|}{0.96} &
  1.00 &
  \multicolumn{1}{c|}{0.93} &
  \multicolumn{1}{c|}{0.9} &
  0.85 &
  \multicolumn{1}{c|}{0.96} &
  \multicolumn{1}{c|}{0.93} &
  0.92 \\ \cline{2-11} 
 &
  \begin{tabular}[c]{@{}c@{}}Macular \\ Degeneration\end{tabular} &
  \multicolumn{1}{c|}{0.95} &
  \multicolumn{1}{c|}{0.92} &
  0.89 &
  \multicolumn{1}{c|}{1.00} &
  \multicolumn{1}{c|}{0.97} &
  1.00 &
  \multicolumn{1}{c|}{0.97} &
  \multicolumn{1}{c|}{0.94} &
  0.94 \\ \hline
\multirow{2}{*}{\begin{tabular}[c]{@{}l@{}}ResNet101\\ +CNN\end{tabular}} &
  \begin{tabular}[c]{@{}c@{}}Healthy \\ Fundus\end{tabular} &
  \multicolumn{1}{c|}{1.00} &
  \multicolumn{1}{c|}{0.98} &
  0.99 &
  \multicolumn{1}{c|}{1.00} &
  \multicolumn{1}{c|}{1.00} &
  0.97 &
  \multicolumn{1}{c|}{1.00} &
  \multicolumn{1}{c|}{0.99} &
  0.99 \\ \cline{2-11} 
 &
  \begin{tabular}[c]{@{}c@{}}Macular \\ Degeneration\end{tabular} &
  \multicolumn{1}{c|}{1.00} &
  \multicolumn{1}{c|}{1.00} &
  0.98 &
  \multicolumn{1}{c|}{1.00} &
  \multicolumn{1}{c|}{0.99} &
  0.99 &
  \multicolumn{1}{c|}{1.00} &
  \multicolumn{1}{c|}{0.99} &
  0.98 \\ \hline
\multirow{2}{*}{\begin{tabular}[c]{@{}l@{}}ResNet101v2\\ +CNN\end{tabular}} &
  \begin{tabular}[c]{@{}c@{}}Healthy \\ Fundus\end{tabular} &
  \multicolumn{1}{c|}{0.96} &
  \multicolumn{1}{c|}{1.00} &
  1.00 &
  \multicolumn{1}{c|}{0.96} &
  \multicolumn{1}{c|}{0.91} &
  0.79 &
  \multicolumn{1}{c|}{0.96} &
  \multicolumn{1}{c|}{0.95} &
  0.89 \\ \cline{2-11} 
 &
  \begin{tabular}[c]{@{}c@{}}Macular \\ Degeneration\end{tabular} &
  \multicolumn{1}{c|}{0.97} &
  \multicolumn{1}{c|}{0.93} &
  0.86 &
  \multicolumn{1}{c|}{0.97} &
  \multicolumn{1}{c|}{1.00} &
  1.00 &
  \multicolumn{1}{c|}{0.97} &
  \multicolumn{1}{c|}{0.97} &
  0.92 \\ \hline
\multirow{2}{*}{\begin{tabular}[c]{@{}l@{}}ResNet152\\ +CNN\end{tabular}} &
  \begin{tabular}[c]{@{}c@{}}Healthy \\ Fundus\end{tabular} &
  \multicolumn{1}{c|}{1.00} &
  \multicolumn{1}{c|}{1.00} &
  0.97 &
  \multicolumn{1}{c|}{1.00} &
  \multicolumn{1}{c|}{0.98} &
  0.96 &
  \multicolumn{1}{c|}{1.00} &
  \multicolumn{1}{c|}{0.99} &
  0.97 \\ \cline{2-11} 
 &
  \begin{tabular}[c]{@{}c@{}}Macular \\ Degeneration\end{tabular} &
  \multicolumn{1}{c|}{1.00} &
  \multicolumn{1}{c|}{0.99} &
  0.97 &
  \multicolumn{1}{c|}{1.00} &
  \multicolumn{1}{c|}{1.00} &
  0.98 &
  \multicolumn{1}{c|}{1.00} &
  \multicolumn{1}{c|}{0.99} &
  0.97 \\ \hline
\multirow{2}{*}{\begin{tabular}[c]{@{}l@{}}ResNet152v2\\ +CNN\end{tabular}} &
  \begin{tabular}[c]{@{}c@{}}Healthy \\ Fundus\end{tabular} &
  \multicolumn{1}{c|}{1.00} &
  \multicolumn{1}{c|}{1.00} &
  0.93 &
  \multicolumn{1}{c|}{0.81} &
  \multicolumn{1}{c|}{0.84} &
  0.92 &
  \multicolumn{1}{c|}{0.90} &
  \multicolumn{1}{c|}{0.92} &
  0.93 \\ \cline{2-11} 
 &
  \begin{tabular}[c]{@{}c@{}}Macular\\  Degeneration\end{tabular} &
  \multicolumn{1}{c|}{0.88} &
  \multicolumn{1}{c|}{0.89} &
  0.94 &
  \multicolumn{1}{c|}{1.00} &
  \multicolumn{1}{c|}{1.00} &
  0.95 &
  \multicolumn{1}{c|}{0.94} &
  \multicolumn{1}{c|}{0.94} &
  0.94 \\ \hline
\end{longtable}

\item Training set is 50\% and Testing set is 50\% where for testing it has used 168 of Macular Degeneration Fundus Images and 136 of Normal Fundus images.
\end{itemize}

\begin{sloppypar}
From the table~\ref{tab1} and table~\ref{tab2} we can see that the leading performed model is ResNet50+CNN with test accuracy of 98.7\% in 90\% train split. Also the accuracy is not decreased if the splitting of train-test is changed. The lowest performed model is ResNet152v2 with accuracy of 85.9\%. The convolutional neural network model is not performed that bad with accuracy of 97\%. But it can say that in convolutional neural when the splitting size is increased the accuracy starts to decrease. On the other hand, ResNet50+CNN, ResNet101+CNN, RestNet152+CNN has performed almost similarly what if the data is split or not. Among these three types of models reason behind choosing RestNet50+CNN as best model because the precision, sensitivity and F1-score of ResNet50+CNN is best. But again you can say normal CNN model is the best model only if you Split the data in 90\% test and 10\% of train.
\end{sloppypar}

\subsection{Confusion Matrix Analysis}
\begin{figure*}[t]
\centering
  \includegraphics[width=10cm, height=12cm]{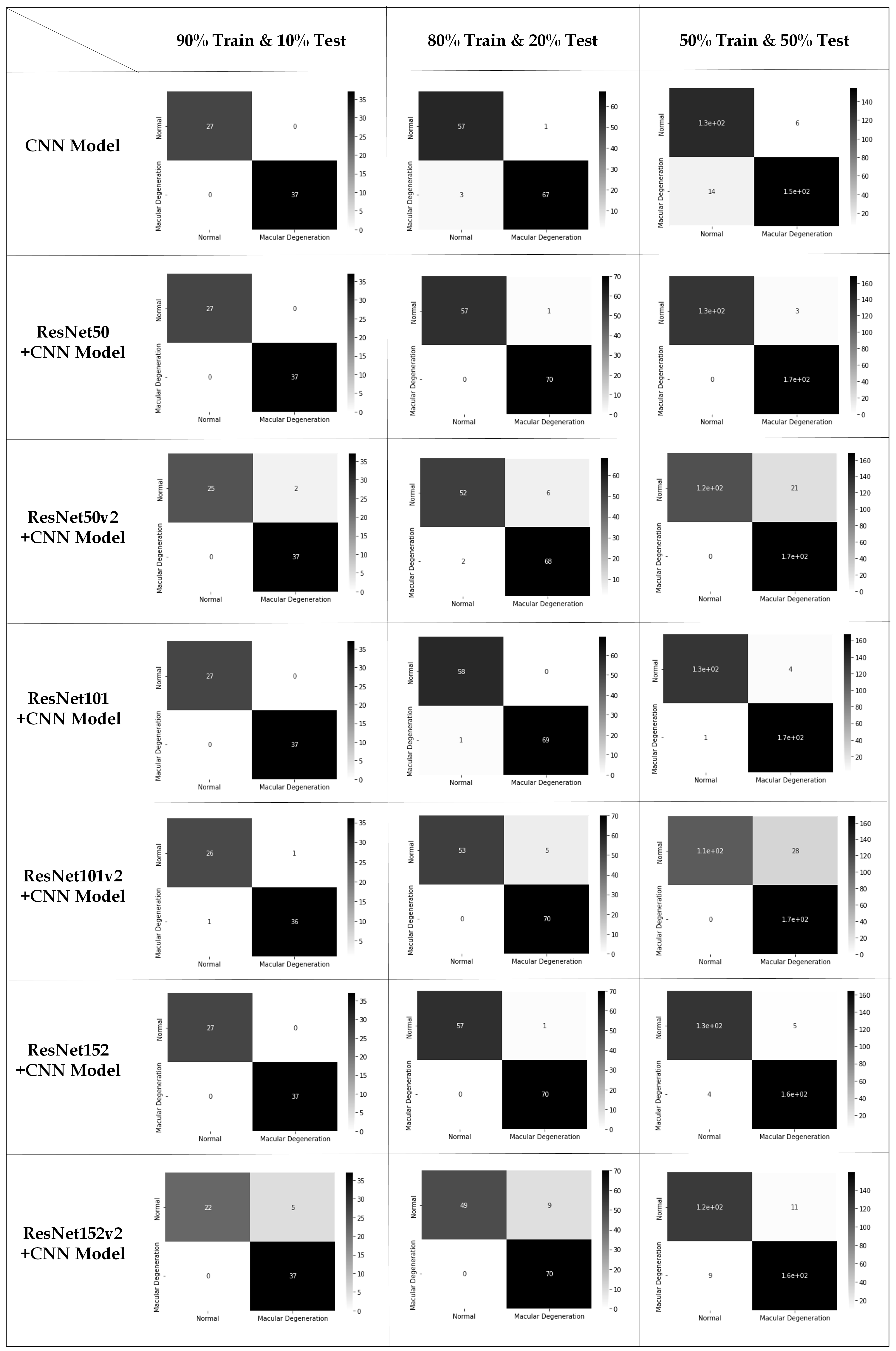}
\caption{Confusion Matrix Analysis}
\label{fig4}       
\end{figure*}

From the confusion matrix, the effectiveness of the model can be determined. The best the effectiveness means the best performances of the model. As from the figure~\ref{fig4} we can see that when we split the data 90\% train and 10\% test, four models (CNN, ResNet50+CNN, ResNet101+CNN, ResNet152+CNN) can easily classify both types of fundus. So, let’s consider those four models as the best model. Then we can see that when we split the data 80\% train and 20\% test, CNN model misclassifies four fundus images. So in this stage the CNN model has be disqualified. Lastly when the dataset is split into 50\% train and 50\% test, ResNet50+CNN model misclassifies the least amount of fundus among the three other models. So according to this analysis we can say that the best model is ResNet50+CNN.

\begin{figure*}[t]
\centering
  \includegraphics[width=5.2cm, height=5.5cm]{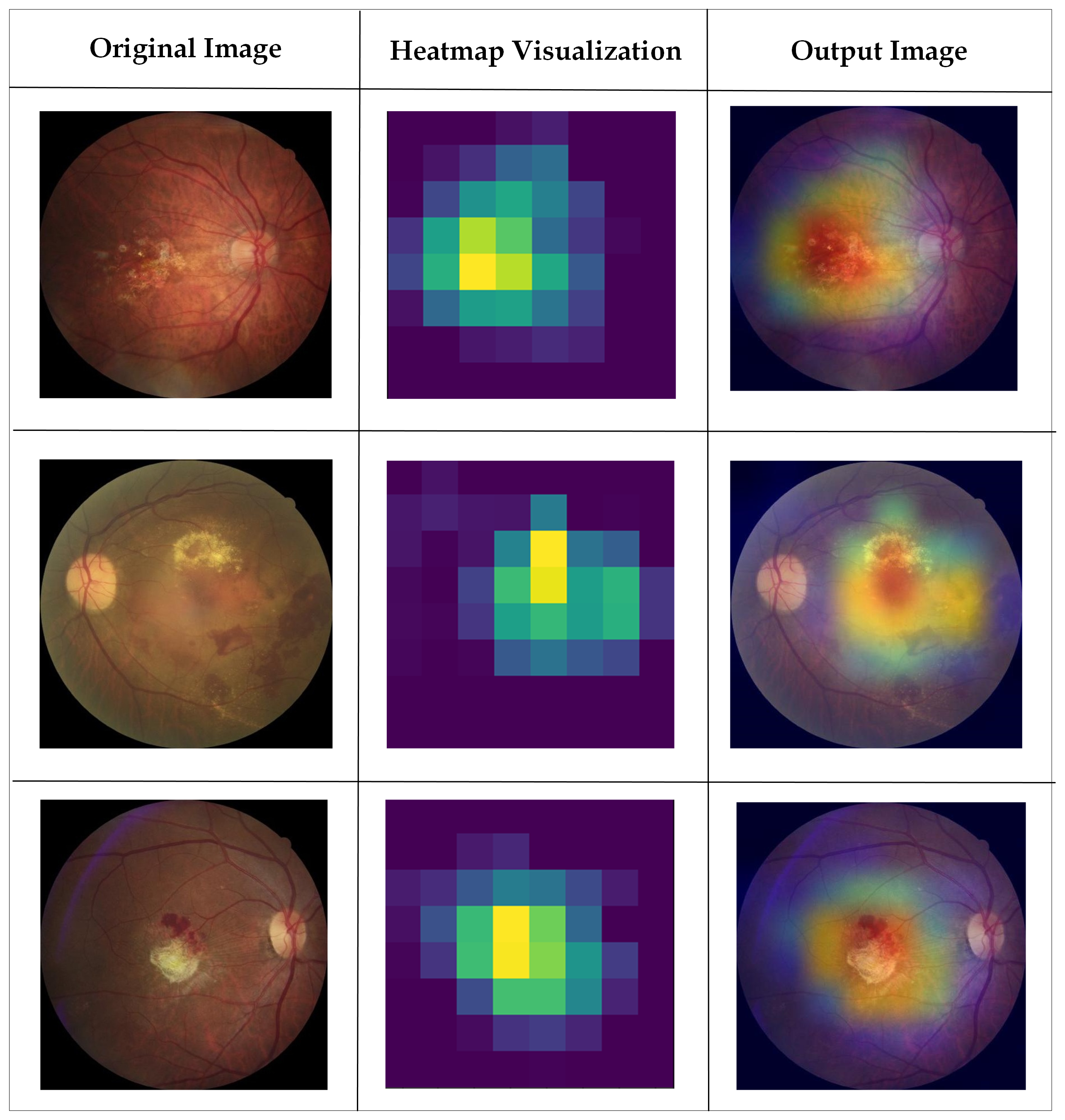}
\caption{The Localization of Macular Degeneration using Grad-CAM Method. Left: Original Input Images. Middle: Heatmap Generated from the input image Right: Superimposed Output Images localizing the effected region.}
\label{fig5}       
\end{figure*}

\subsection{Grad-CAM Analysis}
We have got the best model which is CNN with backbone of ResNet50 model. From the figure~\ref{fig5}. we can see that how the model is working using the Grad-CAM. Firstly, a heatmap has created from the original image to get the important region or the affected area. Then heatmap is applied to the main fundus image to generate the superimposed image which localize the affected area. It will help the practitioner to easily detect the macular degeneration disease fundus and how much area is affected on that fundus. It should be emphasized that the model mainly employs symptoms to identify Macular Degeneration fundus images, and no biases are applied, based on a visual inspection of significant areas.

\section{Conclusion}
The purpose of this study is to discover the optimum model for classifying macular degeneration-affected fundus images into healthy and unhealthy fundus categories. The highest accuracy of 98.7\% was attained using the Convolutional Neural Network with ResNet50 as the backbone. We used the Grad-CAM method to generate a visual depiction of how the model is working on the input images and localized the region of interest areas to test the model's reliability. Even though the dataset is limited our model has shown itself to be capable. Furthermore, the pathologists can readily locate diseases thanks to the employment of the visualization approach, which provides a precise and sharp visualization. In the future, we'll concentrate on classifying more fundus disorders and merging a few datasets.

%

%
%
\bibliographystyle{splncs04}
\bibliography{ref}

\end{document}